\documentclass[conference]{IEEEtran}
\IEEEoverridecommandlockouts
\usepackage{graphicx}
\usepackage{commath}
\usepackage{bm}
\usepackage[T1]{fontenc}
\usepackage{hyperref} 
\usepackage{url}      
\usepackage{booktabs}   
\usepackage{amsfonts} 
\usepackage{nicefrac}    
\usepackage{cite}
\usepackage{color}
\usepackage{array}
\usepackage{mdwmath}
\usepackage{mdwtab}
\usepackage{amssymb,amsmath,amsthm,mathtools}
\usepackage{enumerate}
\usepackage{setspace}
\usepackage{cleveref}
\usepackage{float}
\usepackage[ruled]{algorithm2e}
\usepackage{algorithmic}
\usepackage{textcomp}
\usepackage{xcolor}
\def\BibTeX{{\rm B\kern-.05em{\sc i\kern-.025em b}\kern-.08em
    T\kern-.1667em\lower.7ex\hbox{E}\kern-.125emX}}

\usepackage{flushend}

\begin{document}
\title{Reinforcement Learning for Supply Chain Attacks Against Frequency and Voltage Control}

\author{\IEEEauthorblockN{Amr S. Mohamed, Sumin Lee, and Deepa Kundur}
\IEEEauthorblockA{\textit{The Edward S. Rogers Sr. Department of Electrical \& Computer Engineering} \\
\textit{University of Toronto, Canada}\\
\texttt{\{\{amr.mohamed, suminlee.lee\}@mail, dkundur@ece\}.utoronto.ca}}
% \and
}

\maketitle

\begin{abstract}
The ongoing modernization of the power system, involving new equipment installations and upgrades, exposes the power system to the introduction of malware into its operation through supply chain attacks. Supply chain attacks present a significant threat to power systems, allowing cybercriminals to bypass network defenses and execute deliberate attacks at the physical layer. 
Given the exponential advancements in machine intelligence, cybercriminals will leverage this technology to create sophisticated and adaptable attacks that can be incorporated into supply chain attacks. 
We demonstrate the use of reinforcement learning for developing intelligent attacks incorporated into supply chain attacks against generation control devices. 
We simulate potential disturbances impacting frequency and voltage regulation.
The presented method can provide valuable guidance for defending against supply chain attacks.
\end{abstract}


\IEEEpeerreviewmaketitle

% \linenumbers

\section{Introduction} \label{sec:intro}

The growing resources available to cybercriminals and the financial sponsoring of cyberattacks are producing advanced cyberattacks against industrial control systems. 
Particularly, supply chain attacks pose significant threats to critical infrastructure as they are difficult to protect against.
These attacks involve cybercriminals compromising the supply chain of third-party software and equipment to inject vulnerabilities into devices either before their shipment or through subsequent firmware updates \cite{duman2019modeling}. 
Leveraging third-party equipment makes it challenging for the targeted facility to anticipate and detect the attack. 
Following breach, the malware can spread to other equipment to give the cybercriminals a stronger foothold on the system and/or autonomously disrupt the operation of the infected system or cause damage to the system over a long time.

The potential impact of supply chain attacks is evident in the Stuxnet malware attack against the Iranian Natanz nuclear facility, which leveraged four zero-day exploits and vulnerabilities in Microsoft and Siemens software and equipment \cite{langner2011stuxnet}. 
Stuxnet infected nuclear centrifuges' programmable logic controllers, issuing malicious control commands that caused damage to the centrifuges while hiding its activity to avoid detection \cite{kushner2013real}.

Supply chain attacks can enable cyberattackers to target control loops that are infeasible to compromise remotely. This facilitates the execution of stealthy and damaging attacks.
Demonstrating the impact of compromising physical devices, the Aurora generator test showed that breaching control of the circuit breaker of a generator can enable a cybercriminal to do irreparable damage the generator -- causing significant financial loss to owner-operators and the electric grid \cite{aurorawired}.

The modernization of electric grid infrastructure and replacement of outdated and obsolete equipment is expected to introduce vulnerabilities and provide opportunities for supply chain attacks against the electric grid. 
Given the significant potential for supply chain attacks to disrupt the electric grid, it is essential to understand the attack strategies that might be employed in a supply chain attack. Consequently, we emphasize the need to model intelligent supply chain attacks to study potential physical impacts as a step towards improving the security posture of the power system.

Reinforcement learning (RL) presents a promising method to model and learn intelligent cyber-physical attacks. 
The authors in~\cite{chen2018evaluation} used RL to develop malware that infects substations, falsifying power measurements to compromise state estimation. 
The malware causes voltage sags and subsequent potential cascading failure induced by low-voltage protection generation tripping.
Further studying the use of RL to strategize cyber-physical power system attacks, the authors in~\cite{yan2016q, ni2017reinforcement, paul2018study, ni2019multistage} developed RL agents to synthesize line-switching attacks, which exploit how sudden changes in grid topology can lead to cascading failures and blackout.
Wang \textit{et al.}~\cite{wang2020coordinated} proposed a combined RL-generated line-switching attack involving a physical attack that trips a transmission line and a simultaneous cyberattack that fakes its outage signal on a different line to cause improper dispatch actions.
The above studies only consider the impact of attacks on state estimation and dispatch subsequent to it, without consideration of the dynamic behavior of the electric grid.
Considering attack impact on dynamic frequency regulation, Mohamed \textit{et al.}~\cite{mohamed2023use} developed RL agents to synthesize attacks compromising load-frequency control, including load-switching and false data injection attacks.

In this paper, we apply RL to model intelligent supply chain attacks. 
Our work expands on previous research on RL for power system cyber-physical security by addressing a gap in assessing how cyberattacks targeting voltage regulation can disrupt power system dynamics (rather than state estimation and static power flow). 
Since voltage control is dispatched following optimal power flow, the literature studying voltage attacks has only considered voltage dispatch falsification on a relatively long time-scale (5 minutes per state estimation). 
However, supply chain attacks that involve infecting devices responsible for automatic voltage regulation (AVR), power system stabilization (PSS), or voltage synchronization can execute voltage attacks at a much shorter timescale and exploit the fast voltage regulation dynamics to execute aggressive destabilizing attacks. 
Further, coordinated supply chain attacks -- that might occur following malware propagation or within an advanced persistent threat -- can cause more disruptive impacts. 
Hence, we develop RL-based malware to execute supply chain attacks and demonstrate potential impacts on voltage and frequency regulation and stability. 
Further, we demonstrate the impact of combined supply chain attacks compromising multiple devices simultaneously. 

The paper outline is as follows:
In Section \ref{sec:problem}, we discuss supply chain attacks in the context of compromising generation control devices. 
In Section \ref{sec:method}, we formulate supply chain attacks as a RL problem. In Section \ref{sec:results}, we simulate the impact of several test cases on power system frequency and voltage stability. The conclusion is in Section \ref{sec:conclusion}.

\section{Problem Formulation} \label{sec:problem}

% - AVR and Governor are local IEDs
% - PSS?
% - Difficult to target 
% - Malware introduced into local area network can spread
% - Replacement might be infected
% - USB attacks, disgruntled or unaware employees
% - Malware searches for specific vulnerabilities to be able to rootkit and infect code - vulnerabilities of third-party vendors
% - Infect local device bypassing airgap
% - Strategy to disrupt grid 
% - frequency and voltage fluctuation - what do they cause?
% - tripping worst case
% - PSS sees frequency and gives voltage setpoint
% - Governor sees frequency and gives freq setpoint
% - RL to train a operation which observes the input and devises a malicious output
% - Operation can be uploaded to infected code as part of malware
% - Problem: train RL to define an action policy to learn the impact

\begin{figure}
    \centering
    \includegraphics[width=0.7\columnwidth]{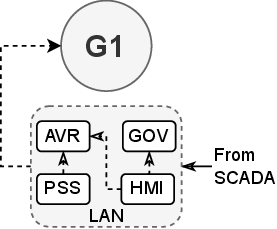}
    \caption{Generator with associated control devices in its LAN. The PSS and AVR regulate the generator's voltage. The governor (GOV) regulates the generator's frequency. The HMI represents the human-machine interface that the owner would operate within the LAN. SCADA dispatch includes automatic generation control and voltage control dispatch.}
    \label{fig:gen}
\end{figure}

Intelligent electronic devices (IED), including the AVR, PSS, and governor (refer to Fig. \ref{fig:gen}), regulate generation to maintain power system stability \cite{tomsovic2005designing}. 
Local to their generation facility, these IEDs communicate within the plant local-area-network (LAN) via Ethernet, acquiring local measurements and sending local messages \cite{sridhar2011cyber}. 
The lack of remote communication to these IEDs makes it largely infeasible for cyberattackers to remotely compromise these IEDs \cite{sridhar2011cyber}.
Nevertheless, these IEDs are vulnerable to cyberattacks that infiltrate the plant LAN, malware that is introduced via infected USB devices \cite{sridhar2011cyber}, or malware that is programmed to the IEDs in a supply chain attack.
Similar to Stuxnet, the malware can be programmed to search for specific software vulnerabilities on the IEDs. Next, the malware can execute a rootkit attack, modifying a portion of the IED program.
To remain hidden, the rootkit can include further specifications to remain dormant, sparsely falsify control to disrupt the grid, or report normal operation values to operators and device logs.

The malware can disrupt the grid by inducing voltage and frequency fluctuations.
Small fluctuations can degrade power quality, cause invisible damage to power system equipment and consumer digital devices over a long period of time, reduce equipment operational life and power system efficiency, and cause flicker \cite{abb2021poor}.
High fluctuations can force equipment tripping, destabilize the power system, and cause cascading failures.

Our research applies RL to develop the attack policy that the malware can upload into the infected IEDs. 
In RL, an agent is trained to learn a policy, mapping a set of observations to actions, to maximize a reward signal in an environment. 
The RL agent's learnt policy can be employed as the malware attack policy that will map the measurement inputs (observations) to the IED to attack actions. The agent is rewarded positively in relation to its disruption of the power system.
The RL's policy involves a mathematical function that can be programmed into the malware. 
For example, the PSS monitors the local frequency for oscillations and sends control signals to the generator AVR to quickly respond to and dampen these oscillations. As illustrated in Fig. \ref{fig:pss}, an RL-based PSS malware would involve mapping the local voltage and frequency measurements to malicious control signals to the AVR.
Similarly, RL-based rootkit infections of the governor (refer to Fig. \ref{fig:gov}) and AVR would involve mapping frequency and voltage observations, respectively, to falsified measurements or malicious control set-points. 

% - Multiple attacks can happen at once 
% - APT can target multiple generation facilities 
% - Use the impact to design physical mitigation strategies
% - Dsiclaimer: preliminary assessment in this paper - future research on mitigation strategies (encourage others to use RL to test their mitigation strategies) and evaluate on more complex systems.
% - Why not completely rewrite the logic? 

We also consider simultaneous supply chain attacks attacks, which might happen following the infection of multiple IEDs at a plant or multiple plants. The malware at one location can remain dormant for a long time until other devices or facilities are infected, and then a simultaneous attack is launched. 

While we do not consider mitigation strategies in this paper, the ultimate goal of modelling attack policies is to guide the design of defense strategies to enhance power system security. 
In this paper, we present a preliminary study to demonstrate the use of RL for devising supply chain attacks and rely on future work to scale the presented method to more complex power systems and develop defenses.

\section{RL for Supply Chain Attacks} \label{sec:method}

\begin{figure}
    \centering
    \includegraphics[width=\columnwidth,trim={0 0cm 0 0cm},clip]{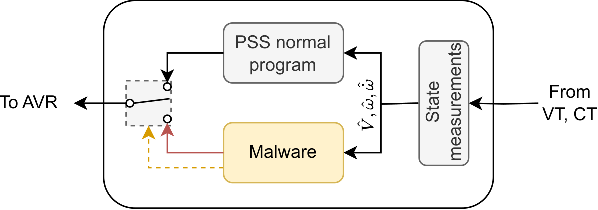}
    \caption{Malware infection of PSS IED. The malware observes the voltage, frequency, and rate-of-change of frequency measurement inputs to the PSS and computes the PSS signal to the AVR. The malware can switch between normal PSS program in-dormancy and its malicious code.}
    \label{fig:pss}
\end{figure}

\begin{figure}
    \centering
    \includegraphics[width=0.7\columnwidth]{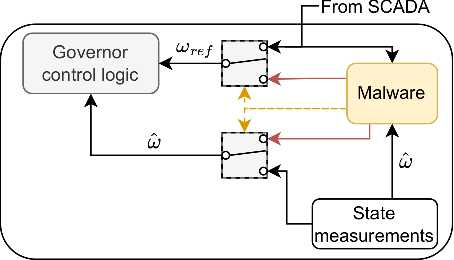}
    \caption{Malware infection of governor IED. The malware observes the  frequency measurement input to the governor (and can additionally estimate the rate-of-change of frequency). The malware can falsify the frequency reference (from SCADA) and/or local frequency measurement fed into the governor control logic.}
    \label{fig:gov}
\end{figure}

% \begin{figure}
%     \centering
%     \includegraphics[width=0.7\columnwidth]{figures/CRL2diagrams-AVR.drawio.eps}
%     \caption{}
%     \label{fig:avr}
% \end{figure}

% - Kundur 2 area system [image]
% - 4 genrou models 
% - exciters and governor models
% - actions space (Vref and wref)
% - tamper with control logic while remaining stealthy to provide opportunity for multiple attack attempts
% - image of just small addition to logic
% - RL PPO architecture
% - actor and critic
% - actor becomes rootkit payload operation

We use the RL Proximal Policy Optimization (PPO) algorithm to model the malware.
The PPO algorithm has a few advantages that makes it suitable for modelling the malware \cite{schulman2017proximal}: First, the algorithm has been shown to achieve state-of-the-art performance on a wide range of continuous control tasks. Likewise, the malware is developed for continuous observations (voltage and frequency measurement inputs to the IEDs) and continuous actions (false control signals). 
Second, in terms of performance, the PPO algorithm enables computationally efficient and stable policy learning.
In this section, we will formulate supply chain attacks as a RL problem.

\begin{figure}
    \centering
    \includegraphics[width=1\columnwidth,trim={1cm 0cm 0.7cm 0cm},clip]{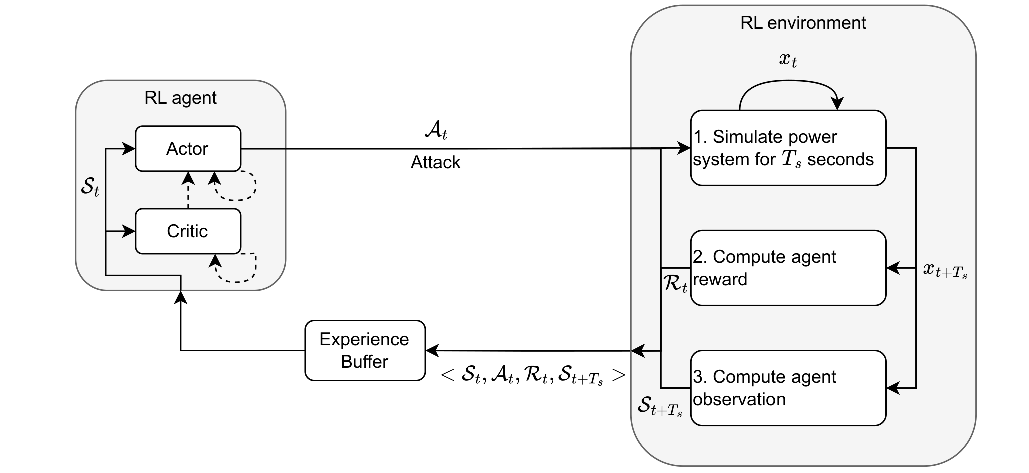}
    \caption{Overview of RL training. The RL environment represents the power system and the RL agent action represents the supply chain attack. The experience buffer stores dynamic trajectories of the power system during the attacks for agent training.}
    \label{fig:overview}
\end{figure}

The training process for RL malware involves iteratively interacting with a model of the power system, which serves as the RL environment, to optimize the malware's attack policy. The process is visually represented in Fig.~\ref{fig:overview}.
The RL agent is composed of an actor, responsible for generating the malware's actions, specifically the false commands or data injections for the supply chain attack. These actions are determined based on the observations of the RL agent. Additionally, the RL agent comprises a critic that evaluates the actions taken by the actor.

During the interaction, the RL agent receives a vector $\mathcal{S}_t$ representing the current state of the power system. Using this information, the actor generates a vector $\mathcal{A}_t$ containing the false data to be injected into the power system's IED device(s).
The RL-based malware can be mathematically represented as a policy
\begin{align}
    \pi(\mathcal{A}_t | \mathcal{S}_t): \mathcal{S}_t \rightarrow \mathcal{A}_t
\end{align}

We simulate the response of the power system to the injected false data for a time-step $T_s$ seconds. The power system can be represented as a non-linear system 
\begin{align}
    \dot{x} &= f(x, \mathcal{A})\\
    \mathcal{S} &= g(x)
\end{align}
with state and output functions $f$ and $g$, respectively. 

Based on the impact of the injected false data on the power system, we return a reward to the RL-based malware. In this study, we reward the agent based on negative impact on the power quality quantified in terms of frequency fluctuation. The general template of the reward function that we use is as follows:
\begin{equation}
    \mathcal{R} = \sum_{g \in \mathcal{G}} \gamma_1 \hat{\dot{\omega}}_g^2 + \gamma_2 \{\texttt{trip}_g\}
\end{equation}
where $\mathcal{G}$ is the set of generators in the power systems and \texttt{trip}$_g$ is a Boolean value, calculated as follows:
\begin{align} \label{eq:trip}
    \texttt{trip}_g &= \{ \hat{V}_g \notin [\underline{V}, \overline{V}] \}
    \vee \{ \hat{\omega}_g \notin [\underline{\omega}, \overline{\omega}] \}
    \vee \{ \abs{\hat{\dot{\omega}}_g} > r \}
\end{align}
signalling when the attack has caused generator $g$ to trip due to the triggering of voltage or frequency protection.
$\hat{V}$, $\hat{\omega}$, and $\hat{\dot{\omega}}$ are the generator's terminal voltage, frequency, and rate-of-change of frequency measurements, respectively. 
The subscript $g$ relates the measurements to generator $g$.
The relay settings corresponding to the different protection functions included in \eqref{eq:trip} are listed in Table \ref{tab:relays}. 

Variables $\gamma_1, \gamma_2 > 0$ in \eqref{eq:trip} are reward scaling values.
We formulate \eqref{eq:trip} to reward the agent in proportion to the magnitude of frequency fluctuations that the agent induces all over the power system to degrade power quality. 
Additionally, the agent receives additional rewards if its actions lead to generation loss that may destabilize the grid.
The reward function can be expanded to include additional attack goals.

\begin{table}
    \centering
    \caption{Frequency and voltage protection relay settings per C37.106-2003 \cite{c37-106ieee} and C37.102-2008 \cite{c37-102ieee}, respectively. Typical rate-of-change of frequency relay setting.}
    \begin{tabular}{c|c|c}
        \toprule
        \textbf{Protection} & \textbf{Parameter} & \textbf{Setting value} \\
        \midrule
        Voltage $\hat{V}$ & $\underline{V}$ & $0.7$ pu\\
        & $\overline{V}$ & $[1.3, 1.5]$ pu\\
        \midrule
        Frequency $\hat{\omega}$ & $\underline{\omega}$ & $57.4$ Hz\\
        & $\overline{\omega}$ & $61.7$ Hz\\
        \midrule
        Rate-of-change  & $r$ & $[0.5, 3]$ Hz/s\\
        of frequency $\hat{\dot{\omega}}$ & & \\
        \bottomrule
    \end{tabular}
    \label{tab:relays}
\end{table}
            
The training happens in episodes during which the training loops between computing an RL action, simulating the power system's response, and rewarding the agent. 
The goal of the PPO algorithm is to optimize the policy to maximize the agent's cumulative reward.
The readers are referred to \cite{schulman2017proximal} for more detail about the PPO algorithm.

If we consider the malware's compromise of a PSS IED, the malware's policy is $\pi(V_{PSS} | \hat{V}, \hat{\omega}, \hat{\dot{\omega}})$, i.e., the malware observes local voltage, frequency, and rate-of-change of frequency measurements that the PSS acquires and computes a false control signal ($V_{PSS}$) that is injected into the AVR of the targeted generator. This is illustrated in Fig. \ref{fig:pss}. The malware appends of piece of malicious code to the program that switches between normal PSS control and malicious control to sporadically disturb power system operation. Compromising the AVR is similar.

When compromising the governor, the RL policy is $\pi(\hat{\omega} | \hat{\omega}, \hat{\dot{\omega}})$ or $\pi(\omega_{ref} | \hat{\omega}, \hat{\dot{\omega}})$. 
The malware observes the frequency measurement acquired by the governor IED and computes its rate-of-change. 
Next, the malware computes a false frequency measurement or reference ($\omega_{ref}$) to the governor control logic, as illustrated in Fig. \ref{fig:gov}.

\section{Case Studies} \label{sec:results}

\begin{figure}
    \centering
    \includegraphics[width=\columnwidth]{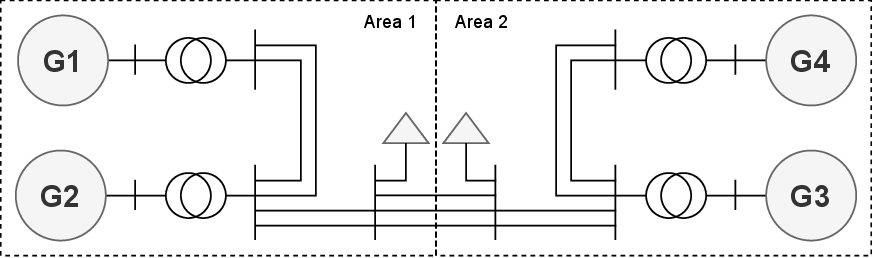}
    \vspace{-6mm}
    \caption{Two area testbed system.}
    \label{fig:2area}
\end{figure}

We use the Kundur two-area system \cite{kundur2022power} (illustrated in Fig. \ref{fig:2area}) in this study.
The two-area system has under-damped modes, which makes it suitable for assessing and demonstrating RL's use for inducing oscillations and compromising power system transient stability in supply chain attacks.

The system contains two coherent groups of generators, each group containing 2 synchronous generators. 
We perform the study using Python. 
We use Andes \cite{andes} to simulate the power system dynamics and package the power system model inside an OpenGym \cite{opengym} environment for training the RL agent. 
The power system parameters can be found in Andes documentation.
The GENROU \cite{genrou}, TGOV \cite{tgov1}, and EXDC2 \cite{exdc2} models are used for the synchronous generators, and their governors and exciters, respectively. 
We use PyTorch \cite{pytorch} and StableBaselines3 \cite{stable-baselines3} for RL. 
Readers can find the code repository for our work \href{https://github.com/amrmsab/RL-CPS-attacks}{here}\footnote{\url{https://github.com/amrmsab/RL-CPS-attacks}}.

We present several test cases below. 
In all case studies, we use a time-step ($T_s$) of 200 milliseconds between the actions of the RL agent. We also average the rate-of-change of frequency values over each 200 milliseconds. 
We train the RL agent in episodes, each 20 seconds long.
The reward scaling values that we apply in the reward function are $\gamma_1 = 1$ s$\cdot$Hz$^{-1}$ and $\gamma_2 = 5$.

The observations spaces of the RL agents in all case studies are limited to the local measurements of the generator(s) that the agent is attacking. For governor IED attacks, the agent can observe the local frequency and its rate of change. For AVR and PSS IED attacks, the agent can additionally observe the voltage measurement.

The RL agents' action spaces are bounded within the relay settings outlined in Table \ref{tab:relays}. We impose this bound to prevent the injection of simple bias attacks that might be easily detected and prevented. Instead, the small bounds encourage the agents to learn more minimal sophisticated attack strategies that induce oscillations in frequency and voltage within the power system with little modification to the compromised signals.

We consider a rate-of-change of frequency relay setting of 1 Hz/s in the studies.
However, we suppress generation tripping during RL training and in the presented case studies to continue to demonstrate the RL attack signal and its impact on the system.

\subsection{Governor IED supply chain attack}

\begin{figure}
    \centering
    \includegraphics[width=\columnwidth]{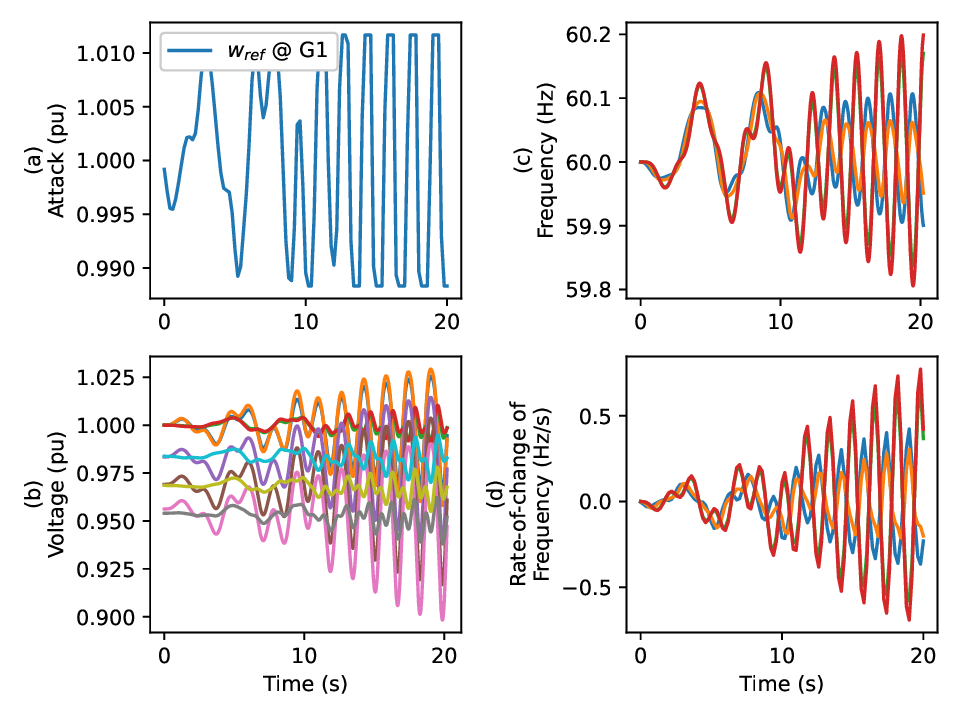}
    \vspace{-6mm}
    \caption{Governor IED supply chain attack. The falsified frequency measurement is in the range $[59.3, 60.7]$ Hz. (a) Falsified frequency measurements injected by attack. (b) Voltage measurements of the 10 buses in the two-area system. (c) Frequency measurements of the 4 generators in the two-area system. (d) Rate of change of frequency of the generators.
    % sampling_time=0.2, episode_length=20,
    % case_file_location='kundur/kundur_full.xlsx',
    % attack_points={1: ['wref0']},
    % observation_points={'v': [], 'omega': [0], 'domega': [0]},
    % freq_bounds_hz=[59.3, 60.7],
    }
    \label{fig:g1wrefsmall}
\end{figure}

In this test case, the malware infects the governor IED of generator G1 and reports false frequency measurements to the governor that are in the range of $[59.3, 60.7]$ Hz. Fig. \ref{fig:g1wrefsmall} shows that the malware's corruption of the frequency measurements introduces frequency fluctuations that grow into approximately $0.6$ Hz/s within 20 seconds (Fig. \ref{fig:g1wrefsmall} (d)). Fig. \ref{fig:g1wrefsmall} (a) shows the reported false frequency measurements and Figs. \ref{fig:g1wrefsmall} (b) and (c) show the impact on bus voltages and generator frequencies, respectively. 

\begin{figure}
    \centering
    \includegraphics[width=\columnwidth]{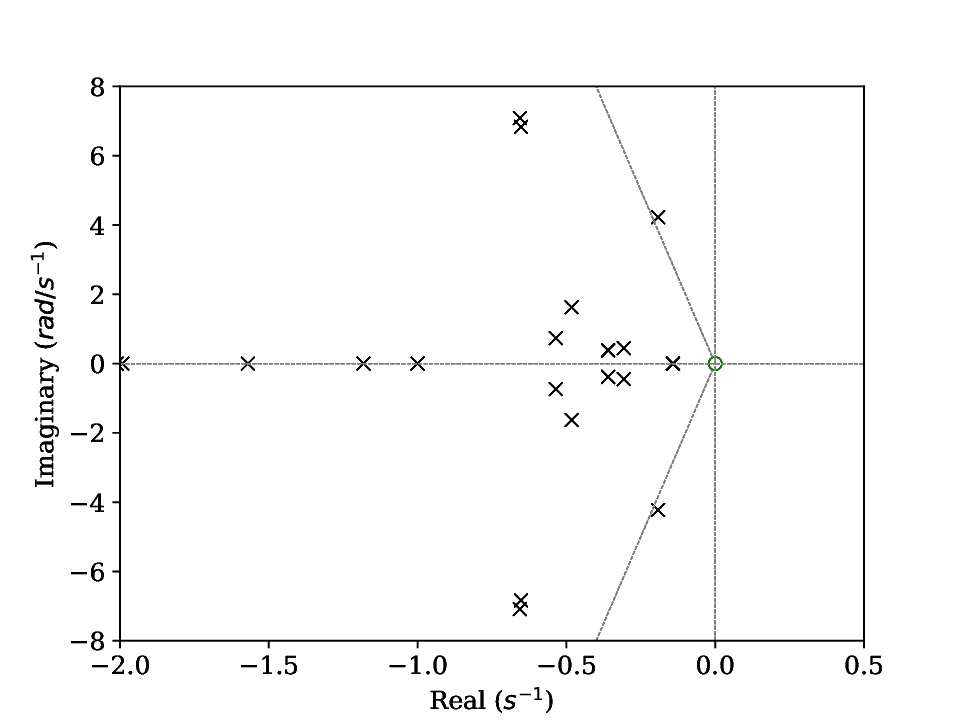}
    \caption{Root locus plot of the eigenmodes of the two area system. The oscillatory eigenmode is located at $-0.20 \pm 4.22j$.
     % -0.19177102+4.2247421j    
    % array([-49.19068854+0.64232883j, -49.19068854-0.64232883j,
    %    -49.20229587+0.39349411j, -49.20229587-0.39349411j,
    %     -0.65640354+7.08595754j,  -0.65640354-7.08595754j,
    %     -0.65252341+6.83425456j,  -0.65252341-6.83425456j,
    %     -0.19177102+4.2247421j ,  -0.19177102-4.2247421j ,
    %     -0.48245324+1.62804574j,  -0.48245324-1.62804574j,
    %     -0.53573049+0.73459147j,  -0.53573049-0.73459147j,
    %     -0.30738804+0.4466424j ,  -0.30738804-0.4466424j ,
    %     -0.36003148+0.38873655j,  -0.36003148-0.38873655j,
    %     -0.35992948+0.37955796j,  -0.35992948-0.37955796j])
    }
    \label{fig:rootlocus}
\end{figure}

\begin{figure}
    \centering
    \includegraphics[width=0.8\columnwidth]{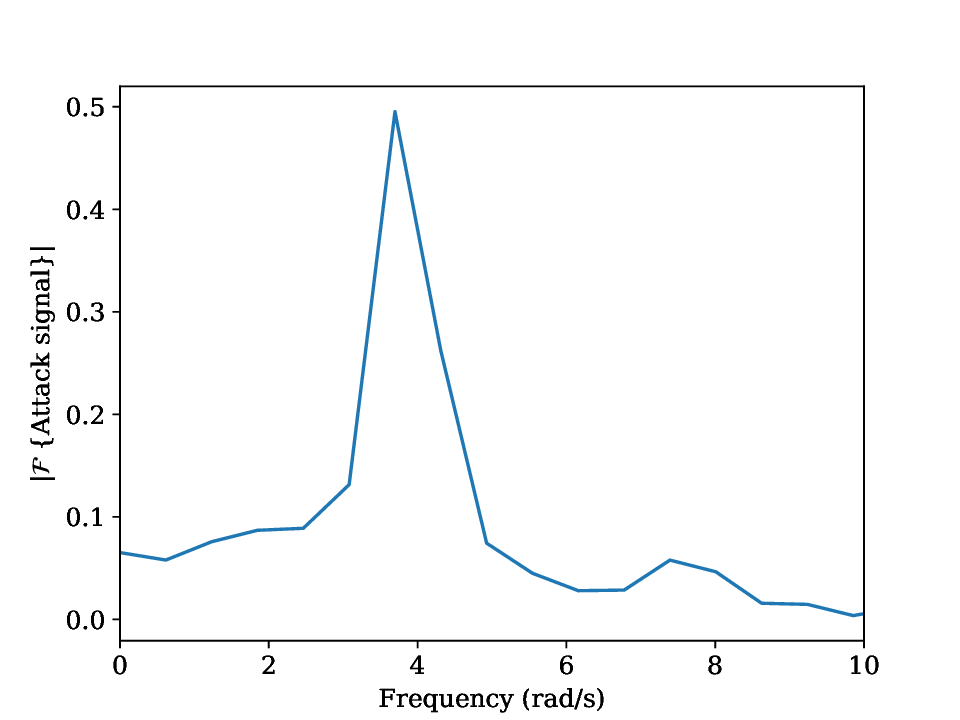}
    \caption{Fourier frequency spectrum of the attack signal in Fig. \ref{fig:g1wrefsmall}. The peak is located at a frequency that is close to the testbed's oscillatory eigenmode.
    % 4.043634108580922 rad/s
    }
    \label{fig:attackfft}
\end{figure}

We observe resonance behavior in the growth of the frequency fluctuations. On further inspection, we find that the RL agent is able to identify and inject an oscillatory attack signal with a frequency that is in close vicinity to the power system's dominant oscillatory eigenmode to excite resonance. 
Fig. \ref{fig:rootlocus} shows the location of the system eigenmodes. The dominant oscillatory eigenmode is located at $4.22$ rad/s.
Fourier analysis of the attack signal, as illustrated in Fig. \ref{fig:attackfft}, shows that the malware injects a signal at $4.04$ rad/s, which excites this eigenmode.

\begin{figure}
    \centering
    \includegraphics[width=\columnwidth]{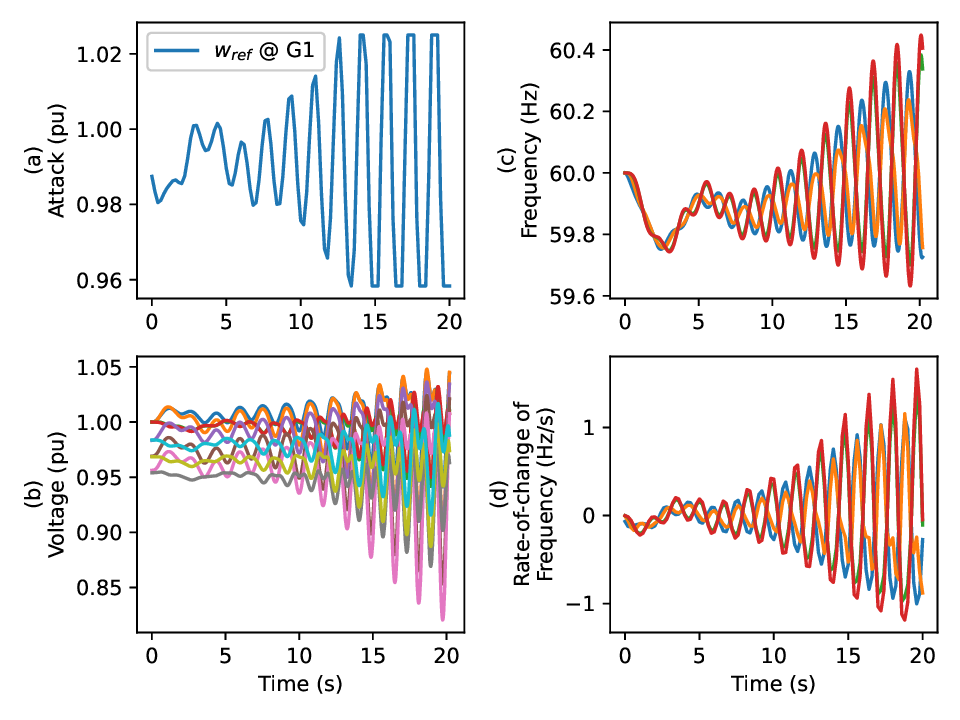}
    \caption{Governor IED supply chain attack. The falsified frequency measurement is in the range $[57.5, 61.5]$ Hz.
    % sampling_time=0.2, episode_length=20,
    % case_file_location='kundur/kundur_full.xlsx',
    % attack_points={1: ['wref0']},
    % observation_points={'v': [], 'omega': [0], 'domega': [0]},
    % freq_bounds_hz=[57.5, 61.5],
    }
    \label{fig:g1wref}
\end{figure}

Note that the attack signal can be easily scaled to induce higher frequency fluctuation. Scaling the reported false frequency measurement to the range of $[57.5, 61.5]$ Hz induces frequency fluctuations that exceed $1$ Hz/s and are very likely to trip rate-of-change of frequency protection in the 2 areas. 
This scaled attack is illustrated in Fig. \ref{fig:g1wref}.

Note that while the attack targets G1 in Area 1, the frequency fluctuations in Area 2 (green and red in Fig. \ref{fig:g1wref} (d)) grow faster than in Area 1, and hence, generators in Area 2 (G3 and G4) would trip sooner than generators in Area 1.
Generation tripping with a rate-of-change of frequency protection setting of 1 Hz/s would happen in less than 15 seconds.
This observation highlights the interconnected nature of the power system, wherein an attack on a generator has the potential to trigger failures in other areas of the system.

\begin{figure}
    \centering
    \includegraphics[width=\columnwidth]{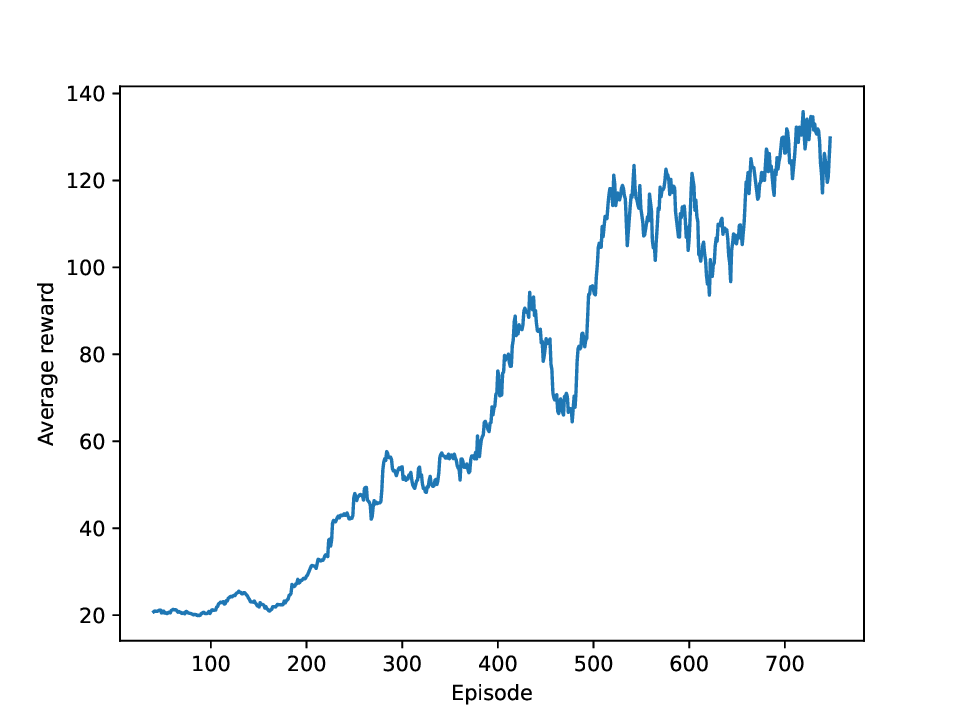}
    \caption{Learning curve for agent in Fig. \ref{fig:g1wref}. The reward is averaged over 40 episodes.
    }
    \label{fig:learn}
\end{figure}

Figure \ref{fig:learn} presents the learning curve of the RL agent depicted in Figure \ref{fig:g1wref}. The plot illustrates the episode reward obtained by the agent during the training process (averaged over 40 episodes for smoothness). The observed growth in the curve is attributed to the agent's learning of policies that result in an increased rate-of-change of frequency, surpassing the rate-of-change of frequency protection setting.

\subsection{Combined governor IED supply chain attack}

\begin{figure}
    \centering
    \includegraphics[width=\columnwidth]{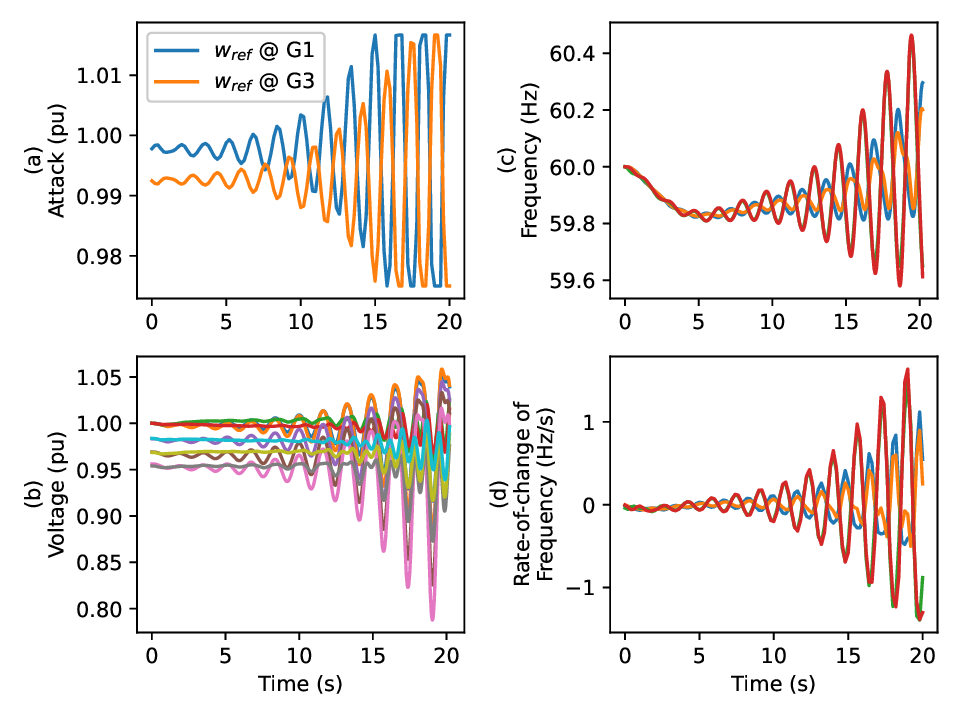}
    \caption{Combined governor IED supply chain attacks. The falsified frequency measurements are in the range $[58.5, 61]$ Hz. 
    % 'sampling_time': 0.2, 'episode_length': 20,
    % 'case_file_location': 'kundur/kundur_full.xlsx',
    % 'attack_points': {1: ['wref0'], 3: ['wref0']},
    % 'observation_points': {'v': [], 'omega': [0, 2], 'domega': [0, 2]},
    % 'freq_bounds_hz': [58.5, 61],
    % 'voltage_bounds_pu': [-0.15, 0.05], },
    }
    \label{fig:g1g3wref}
\end{figure}

Combined supply chain attacks can produce more minimal (in terms of smaller range of reported false frequency measurements), yet more aggressive attacks (in terms of frequency fluctuation).
In Fig. \ref{fig:g1g3wref}, we consider the case when the attack has infected the governor IEDs of generators G1 and G3. The reported false frequency measurements to the governors are limited to the range of $[58.5, 61]$ Hz, which is smaller than the range considered in the test in Fig. \ref{fig:g1wref} but with comparable effects.

\subsection{PSS-governor IED supply chain attack}

\begin{figure}
    \centering
    \includegraphics[width=\columnwidth]{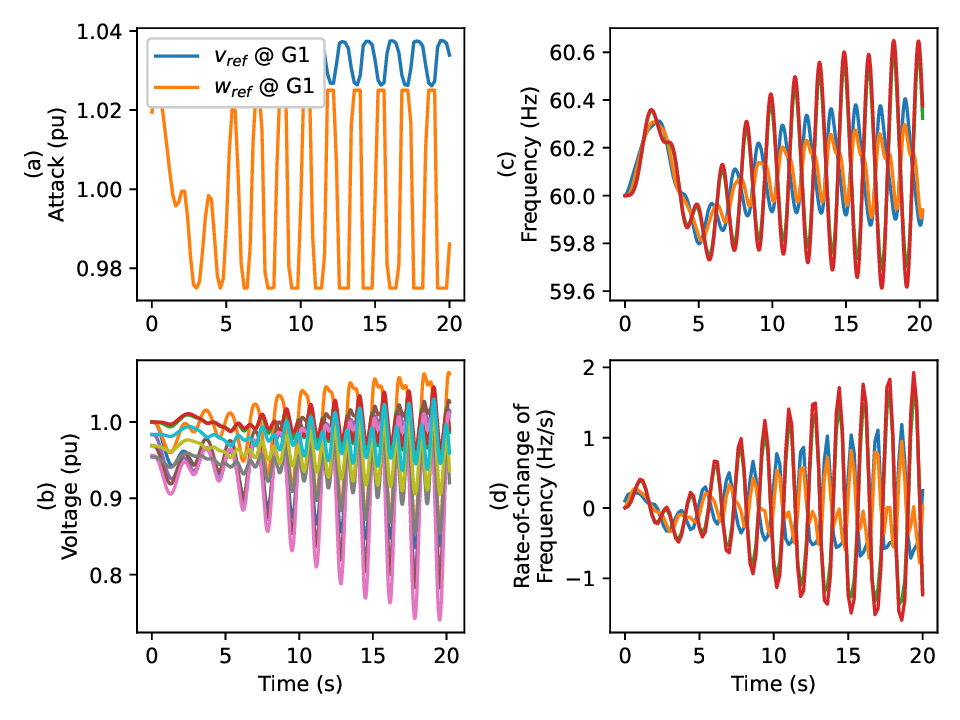}
    \caption{Combined governor and PSS IED supply chain attacks. The falsified frequency measurement is in the range $[58.5, 61]$ Hz. The falsified voltage measurement is in the range $[0.95, 1.12]$ pu.
    % sampling_time=0.2, episode_length=20,
    %         case_file_location='kundur/kundur_full.xlsx',
    %         attack_points={1: ['vref0', 'wref0']},
    %         observation_points={'v': [0], 'omega': [0], 'domega': [0]},
    %         freq_bounds_hz=[58.5, 61.5],
    %         voltage_bounds_pu=[-0.144, 0.026], $[1.026, 1.038]$
    }
    \label{fig:g1vrefwref}
\end{figure}

Alternatively, a combined supply chain attacks against the PSS and governor of one generator (G1) can produce higher frequency fluctuations while also allowing a smaller range of reported false frequency measurements.
In Fig. \ref{fig:g1vrefwref}, the malware reports frequency measurements to the governor in the range of $[58.5, 61.5]$ Hz and voltage measurements to the AVR in the range of $[0.95, 1.12]$ pu.
This amplifies the frequency fluctuations to close to $2$ Hz/s.

\subsection{AVR IED supply chain attack}

\begin{figure}
    \centering
    \includegraphics[width=\columnwidth]{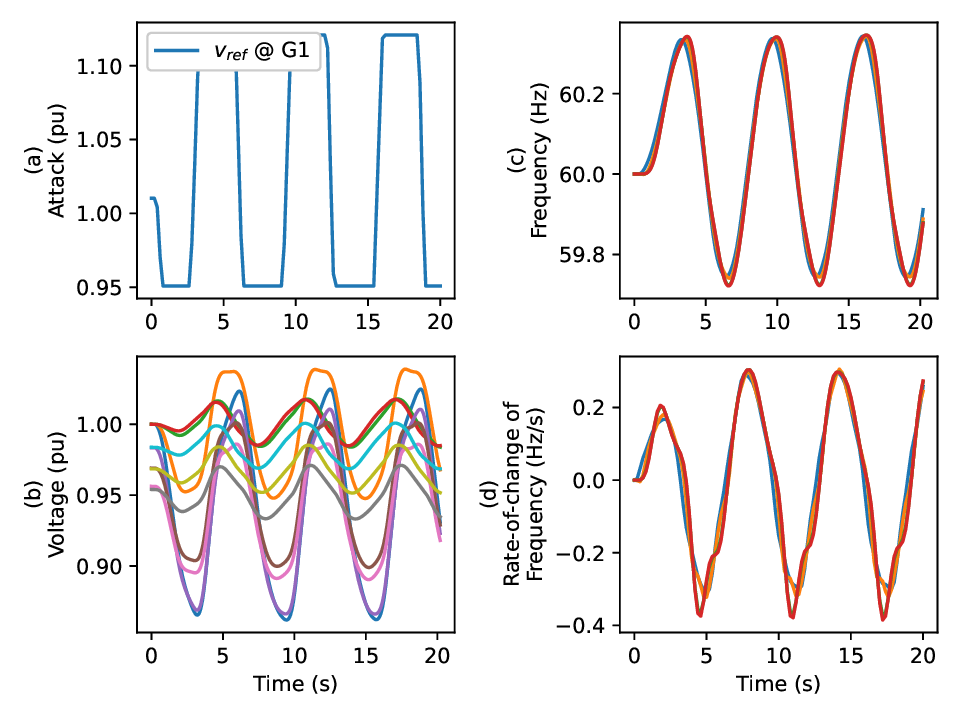}
    \caption{PSS IED supply chain attack. The falsified voltage measurement is in the range $[0.95, 1.15]$ pu.
    % 'sampling_time': 0.2, 'episode_length': 20,
    % 'case_file_location': 'kundur/kundur_full.xlsx',
    % 'attack_points': {1: ['vref0']},
    % 'observation_points': {'v': [], 'omega': [0], 'domega': [0]},
    % 'voltage_bounds_pu': [-0.144, 0.026], },
    }
    \label{fig:g1vref}
\end{figure}

In this test case, the malware infects the AVR (or PSS) IED of generator G1. The malware reports voltage measurements to the AVR that are in the range of $[0.95, 1.15]$ pu. Fig. \ref{fig:g1vref} shows that the malware's corruption of the voltage measurements introduces frequency fluctuations that grow into approximately $0.4$ Hz/s (Fig.~\ref{fig:g1vref} (d)). 

\begin{figure}
    \centering
    \includegraphics[width=\columnwidth]{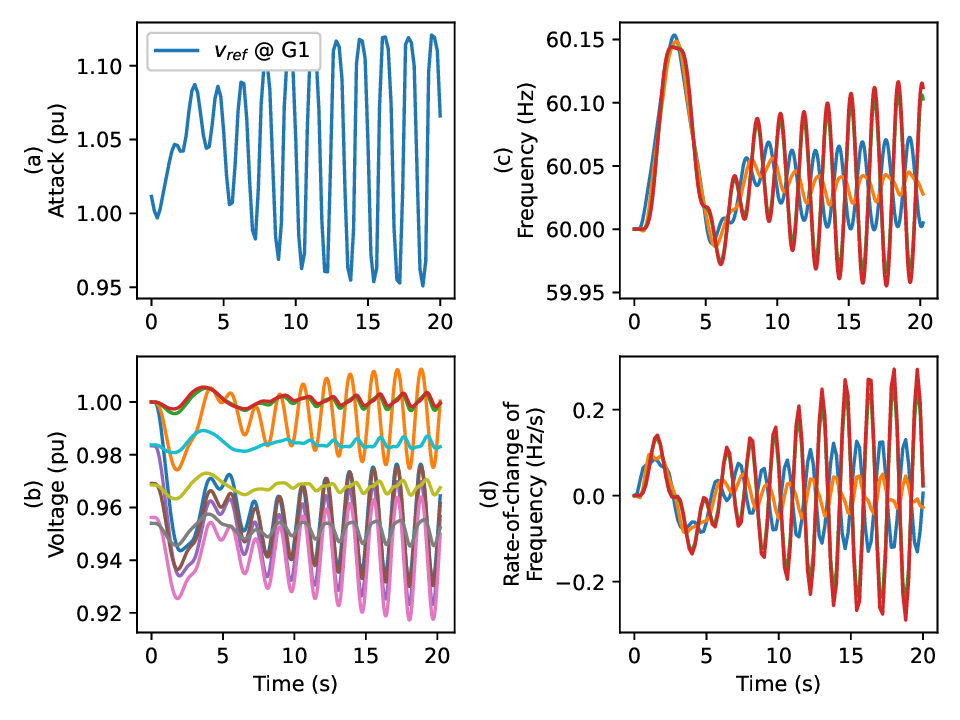}
    \caption{PSS IED supply chain attack. The falsified voltage measurement is in the range $[0.95, 1.15]$ pu. The attack frequency is near the system's oscillatory eigenmode.
    % 'sampling_time': 0.2, 'episode_length': 20,
    % 'case_file_location': 'kundur/kundur_full.xlsx',
    % 'attack_points': {1: ['vref0']},
    % 'observation_points': {'v': [], 'omega': [0], 'domega': [0]},
    % 'voltage_bounds_pu': [-0.144, 0.026], },
    }
    \label{fig:g1vreffast}
\end{figure}

We notice that the attack does not excite the oscillatory eigenmode of the system like previous attacks. The RL training learns that injecting the attack in Fig.~\ref{fig:g1vref} (a) causes larger frequency fluctuation. For comparison, Fig. \ref{fig:g1vreffast} shows a voltage corruption attack that aims to excite the oscillatory eigenmode similar to the previous test cases. The resulting frequency fluctuations are smaller in Fig. \ref{fig:g1vreffast} compared to Fig. \ref{fig:g1vref}. 

% AVR IED supply chain attacks can combine to cause larger frequency fluctuations. 
% Fig. \ref{fig:g1g3vref} illustrates a combined attack infecting the AVR IEDs of generators G1 and G3 that causes more amplified frequency fluctuation exceeding $0.5$ Hz/s.

\subsection{Combined AVR IED supply chain attack}

\begin{figure}
    \centering
    \includegraphics[width=\columnwidth]{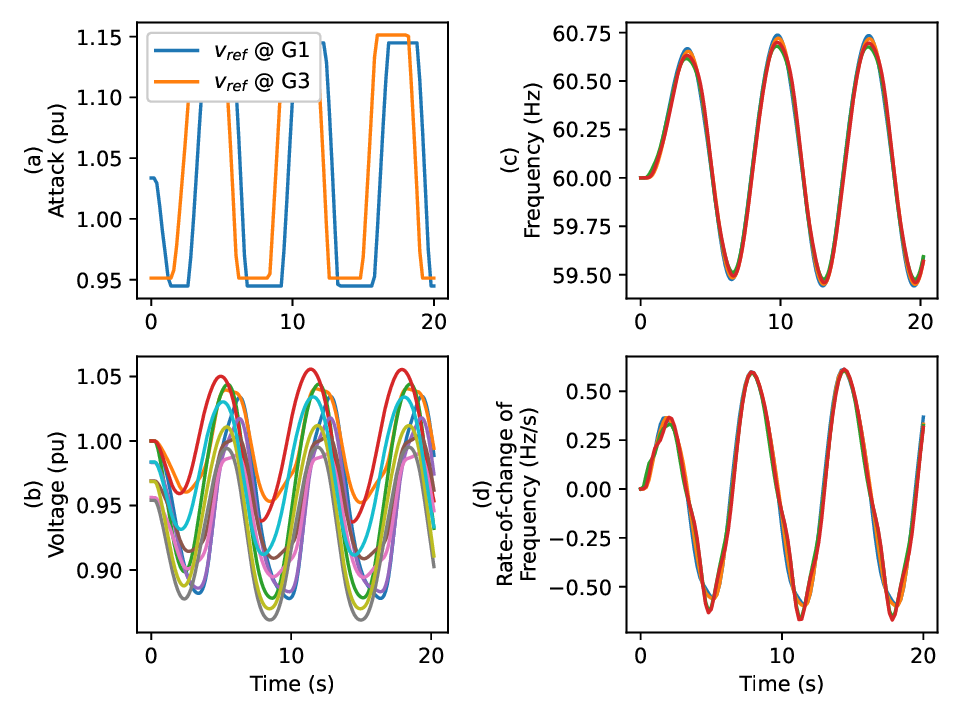}
    \vspace{-7mm}
    \caption{Combined PSS IED supply chain attacks.
    % 'sampling_time': 0.2, 'episode_length': 20,
    % 'case_file_location': 'kundur/kundur_full.xlsx',
    % 'attack_points': {1: ['vref0'], 3: ['vref0']},
    % 'observation_points': {'v': [0,2], 'omega': [0,2], 'domega': [0,2]},
    % 'freq_bounds_hz': [58.5, 61.5],
    % 'voltage_bounds_pu': [-0.15, 0.05], },
    }
    \label{fig:g1g3vref}
\end{figure}

AVR (or PSS) IED supply chain attacks can also combine and lead to more significant frequency fluctuations. 
In Figure \ref{fig:g1g3vref}, we explore the scenario where malware infects the AVR IEDs of generators G1 and G3. 
The reported voltage values to the AVR of G1 are in the range $[0.95, 1.15]$ pu, as considered in Figure \ref{fig:g1vref}. The AVR of G3 receives reported values of $[0.94, 1.14]$ pu. 
In Figure \ref{fig:g1g3vref}, we observe that this combined attack can induce amplified frequency fluctuations that exceed $0.5$ Hz/s.

\section{Conclusion} \label{sec:conclusion}

In this study, we leveraged RL to devise supply chain attacks that compromise local generation control devices, specifically targeting governor, AVR, and PSS IEDs. The results of our research demonstrate the potential of these attacks to degrade power quality and inflict long-term invisible impacts on power system equipment. Further, these attacks can also pose a serious threat to power system stability by forcing generation tripping.

Through several case studies, we have illustrated that RL agents can successfully learn and deploy sophisticated attack policies, including simultaneous attacks. These attack policies can then be packaged into malware, which can be maliciously uploaded to generation control IEDs during supply chain attacks, either before their installation or through system updates.

The implications of our findings underscore the need to employ RL-based approaches for anticipating intelligent supply chain attacks. By proactively employing RL-based defense mechanisms, we can effectively safeguard against the emerging threat landscape and mitigate potential disruptions to power systems.

In conclusion, our research serves as a persuasive call-to-action for the adoption of RL techniques to anticipate and defend against intelligent supply chain attacks.

\appendix

\begin{table}[H]
    \footnotesize
    \centering
    \caption{RL Agent Hyper-parameters}
    \begin{tabular}{l|l}
        \toprule
        \textbf{Parameter} & \textbf{Value}\\
        \midrule
        Learning rate & $3 \times 10^{-4}$\\
        Batch size & $64$\\
        Discount factor & $0.99$\\
        Bias vs variance trade-off factor & $0.95$\\
        Entropy coefficient for loss calculation & $0$\\
        Value function coefficient for loss calculation & $0.5$\\
        Maximum value for gradient clipping & $0.5$\\
        \midrule
        Actor network & Linear($4$, $64$)\\
        & Tanh()\\
        & Linear($64$, $64$)\\
        & Tanh()\\
        & Linear($64$, 2)\\
        \midrule
        Critic network & Linear($4$, $64$)\\
        & Tanh()\\
        & Linear($64$, $64$)\\
        & Tanh()\\
        & Linear($64$, $1$)\\
        \bottomrule
    \end{tabular}
    \label{tab:ppoparameters}
\end{table}

\bibliographystyle{IEEEtran}
\bibliography{refs}

\end{document}